\documentclass{elsarticle}
\usepackage{hyperref}
\usepackage{algorithm}
\usepackage{algpseudocode}

\newcommand{\depend}[1]{\mathit{depend}(#1)}

\begin{document}

\title{Web of Lossy Adapters for Interface Interoperability: An Algorithm and NP-completeness of Minimization}
\author{Yoo Chung}
\ead{chungyc@icu.ac.kr}
\author{Dongman Lee\corref{cor}}
\ead{dlee@cs.kaist.ac.kr}
\address{KAIST ICC, 119 Munjiro, Yuseong-gu, Daejeon 305-732, South Korea}
\cortext[cor]{Corresponding author}

\begin{abstract}
  By using different interface adapters for different methods, it is possible to construct a maximally covering web of interface adapters which incurs minimum loss during interface adaptation.  We introduce a polynomial-time algorithm that can achieve this.  However, we also show that minimizing the number of adapters included in a maximally covering web of interface adapters is an NP-complete problem.
\end{abstract}

\begin{keyword}
  software engineering, interface adaptation, interoperability
\end{keyword}

\journal{Information Processing Letters}

\maketitle

\section{Introduction}
\label{sec:introduction}

Different services that provide similar functionality will often be accessed using widely different interfaces, especially if standardization is lacking.  To avoid having to rewrite separate code for all interfaces that may have to be used, interface adapters can be used to translate calls to one interface into calls to another interface.~\cite{h:gamma1994}

These interface adapters may not be able to achieve perfect adaptation.  It becomes harder to analyze the adaptation loss when combining such interface adapters in order to reduce the number of adapters that must be developed.  Our previous work has defined a rigorous mathematical basis for analyzing the loss in single interface adapter chains.~\cite{chung2009}

This paper extends our previous work by considering the use of a \emph{web} of interface adapters.  This allows the adaptation of interfaces with minimum loss, and we describe a polynomial-time algorithm that can achieve this.  However, we also show that finding a web of adapters that can achieve minimum adaptation loss with the minimum number of interface adapters is an NP-complete problem, which implies that reducing the number of adapters included in a web of interface adapters should be done heuristically.

\section{Preliminaries}
\label{sec:preliminaries}

The basic concepts and notations we use follow those in our previous work.~\cite{chung2009}   We will be using a range convention for the index notation used to express matrixes and vectors.~\cite{index-notation}

We take the view that an interface defines multiple methods, and that an interface adapter converts a call to one method in a source interface into calls to one or more methods in a target interface.  We assume that if a method in a source interface can be adapted, then it can be adapted perfectly.  Any loss would be incurred if a method in a source interface cannot be adapted at all.  We also assume that interface adapters do not store any state.

An \emph{interface adapter graph} is a directed graph where interfaces are nodes and adapters are edges.  If there are interfaces $I_1$ and $I_2$ with an adapter $A$ that adapts source interface $I_1$ to target interface $I_2$, then $I_1$ and $I_2$ would be nodes in the interface adapter graph while $A$ would be a directed edge from $I_1$ to $I_2$.

The \emph{method dependency matrix} $a_{ji}$ for an adapter~$A$ is defined by how the adapter depends on the availability of a method in the source interface in order to implement a method in the target interface.  $a_{ji}$ is true if and only if method~$j$ in the target interface can be implemented only if method~$i$ in the source interface is available.  We denote the method dependency matrix associated with an adapter~$A$ as $\depend{A}$.

\section{Web of lossy adapters}
\label{sec:web-adapters}

Existing approaches use only a single interface adapter to adapt a given target interface from one or more source interfaces.~\cite{gschwind:sem2002,ponnekanti:percom2003,kim:percom2008}  These approaches force us to choose among imperfect chains of interface adapters, where one chain might be able to adapt certain methods in the source interface but cannot adapt other methods covered by another chain and vice versa.  However, using a \emph{web of interface adapters}, a directed acyclic interface adapter graph where different adapters can be used to adapt different methods in an interface, can cover all methods that can possibly be adapted, incurring minimum loss.

\begin{algorithm*}
  \caption{Constructing maximally covering web of interface adapters.}
  \label{alg:maximal-cover}
  \begin{algorithmic}
    \Function{Maximal-Cover}{$G$, $s$, $t$}
      \State $(Q, D, S, M, C) \gets \mbox{\textsc{Cover-Setup}(G, s)}$
      \While{$Q$ is not empty}
        \State extract $(I, i)$ from $Q$
        \For{$(A = (I, I'), j) \in M[I][i]$}
          \If{$C[A][j] > 0$}
            \State $C[A][j] \gets C[A][j] - 1$
            \If{$C[A][j] = 0$}
              \Comment adaptation viable
              \State $D[I'][j] \gets D[I'][j] \cup \{ A \}$
              \If{not $S[I'][j]$}
                \State $S[I'][j] \gets \mathit{true}$
                \State insert $(I', j)$ into $Q$
                \Comment trigger new dependent
              \EndIf
            \EndIf
          \EndIf
        \EndFor
      \EndWhile
      \State \textbf{return} (\textsc{Cover-Subgraph}($D$, $t$), $D$)
    \EndFunction
  \end{algorithmic}
\end{algorithm*}

Algorithm~\ref{alg:maximal-cover} can construct a web of interface adapters that can cover all possible methods in a target interface given a fully functional source interface, which we will refer to as a \emph{maximally covering} web of interface adapters.  It is based on unit propagation for Horn formulae~\cite{h:dowling:jlp1984}, targeted towards building a web of interface adapters.  It works in two phases, where it first computes all methods in all interfaces that can be adapted given the source interface, and then extracts only the subgraph relevant for the target interface.  Algorithms \ref{alg:cover-setup} and \ref{alg:cover-subgraph} are subalgorithms responsible for setup and subgraph extraction, respectively.

\begin{algorithm*}
  \caption{Setup for constructing maximal covering.}
  \label{alg:cover-setup}
  \begin{algorithmic}
    \Function{Cover-Setup}{$G = (V, E)$, $s$}
      \State $Q \gets \mbox{empty queue}$
      \For{$I \in V$ and method $i$ of $I$}
        \State $D[I][i] \gets \emptyset$
        \Comment list of viable adapters
        \State $S[I][i] \gets \mathit{false}$
        \Comment whether satisfiable
        \State $M[I][i] \gets \emptyset$
        \For{$A = (I, I') \in E$}
          \State $M[I][i] \gets M[I][i] \cup \{ (A,j) \,|\, \depend{A}_{ji} \}$
          \Comment dependents
        \EndFor
      \EndFor
      \For{$A = (I_1, I_2) \in E$ and method $j$ of $I_2$}
        \State $C[A][j] \gets |\{i \,|\, \depend{A}_{ji} \}|$
        \Comment unsatisfied dependency count
      \EndFor
      \For{each method $i$ of $s$}
        \Comment start with source interface
        \State $S[s][i] = \mathit{true}$
        \State insert $(s, i)$ into $Q$
      \EndFor
      \State \textbf{return} $(Q, D, S, M, C)$
    \EndFunction
  \end{algorithmic}
\end{algorithm*}

\begin{algorithm}
  \caption{Extract subgraph comprising web of interface adapters.}
  \label{alg:cover-subgraph}
  \begin{algorithmic}
    \Function{Cover-Subgraph}{$D$, $t$}
      \State $V' \gets \emptyset$, $E' \gets \emptyset$
      \State $Q \gets \mbox{empty queue}$, $Q' \gets \emptyset$
      \For{method $i$ of $t$}
        \State insert $(t, i)$ in $Q$ and $Q'$
      \EndFor
      \While{$Q$ is not empty}
        \State extract $(I', j)$ from $Q$
       \State $V' \gets V' \cup \{ I' \}$
        \State $E' \gets E' \cup D[I'][j]$
        \For{$A = (I, I') \in D[I'][j]$}
          \For{$i$ such that $\depend{A}_{ji}$}
            \If{$(I, i) \not\in Q'$}
              \State insert $(I, i)$ into $Q$ and $Q'$
            \EndIf
         \EndFor
        \EndFor
      \EndWhile
      \State \textbf{return} $(V', E')$
    \EndFunction
 \end{algorithmic}
\end{algorithm}

Simply constructing a web of interface adapters is not the goal by itself, of course.  The real goal is to use the interface adapters to adapt methods from a source interface into those of a target interface.  Choosing which adapters should be used for which methods is more complex than in the case for a single chain, where there is no choice at all.  Algorithm~\ref{alg:adapt-method} is an abstract algorithm for determining which interface adapters should be invoked when adapting each method.  It needs more information than just the web of interface adapters, which is provided by the value $D$ also returned in algorithm~\ref{alg:maximal-cover}.

The interface adapters used to adapt a given method are specified by algorithm~\ref{alg:adapt-method}; the concrete steps involved in actually adapting a method are left to how interface adaptation is actually done, whether it be direct invocation by the interface adapter, call substitution after constructing the call graph, or composition of interface adapters specified in a high-level language.  The exact criterion for selecting an adapter in algorithm~\ref{alg:adapt-method} also does not affect the correctness of the algorithm.

\begin{algorithm}
  \caption{Adapting a specific method in the target interface.}
  \label{alg:adapt-method}
  \begin{algorithmic}
    \Procedure{Adapt-Method}{$D$, $s$, $t$, $m$}
      \If{$D[t][m] = \emptyset$}
        \State adaptation not possible
      \Else
        \State \textsc{Recursive-Adapt}($D$, $s$, $t$, $m$, $\{t\}$)
      \EndIf
    \EndProcedure

    \Procedure{Recursive-Adapt}{$D$, $s$, $I$, $m$, $V$}
      \If{$I=s$}
        \State \textbf{return}
      \EndIf
      \State select $A = (I', I) \in D[I][m]$ where $I' \not\in V$
      \For{method $i$ in $I'$ where $\depend{A}_{mi}$}
        \State \textsc{Recursive-Adapt}($D$, $s$, $I'$, $i$, $V \cup \{ I' \}$)
      \EndFor
      \State adapt method $m$ in interface $I$ using $A$
    \EndProcedure
  \end{algorithmic}
\end{algorithm}

Algorithm~\ref{alg:maximal-cover} constructs a maximally covering web of adapters, but it completely ignores the number of interface adapters it incorporates in the web.  It could end up constructing a web with hundreds of interface adapters when less than a dozen would do.  However, trying to minimize the number of incorporated interface adapters is an NP-complete problem as we will show in section~\ref{sec:minimization}.  Invoking the minimum number of interface adapters while actually adapting a method also turns out to be NP-complete.

\section{Minimizing number of adapters}
\label{sec:minimization}

While algorithm~\ref{alg:maximal-cover} can construct a maximally covering web of interface adapters in polynomial time ($O(m^2)$ being a loose time bound with a straightforward implementation, where $m$ is the total number of methods), it is unlikely there will be a polynomial-time algorithm for finding a maximally covering web of adapters with the minimum number of interface adapters.  This is because the problem is NP-complete, which we will prove with a reduction from one-in-three 3SAT.~\cite{h:garey1979}

We formally define MINWEB as the problem of whether there is a web of interface adapters in an interface adapter graph from a given source interface to a given target interface such that it is maximally covering and has at most $K$ interface adapters.  Given a candidate boolean expression for one-in-three 3SAT with $c$~clauses and $v$~variables, we will reduce it to a candidate interface adapter graph for MINWEB such that the boolean expression is an instance of one-in-three 3SAT if and only if there is a maximally covering web of interface adapters with at most $v+2c$ adapters.

For each variable, we create an interface with methods corresponding to all the literals, two for each variable.  For each clause, we create an interface with only a single method.  We also separately create a source interface with methods corresponding to the possible literals and a target interface with methods corresponding to the clauses.

Starting from the source interface, we connect the interfaces corresponding to variables serially.  Between each of these interfaces, we define two adapters, one which makes the method corresponding the successor variable true and the other which makes it false, by making the method correspoding to the positive literal available and the method corresponding to the negative literal unavailable in one adapter and the opposite in the other adapter.  Other literals are left alone.  This is identical to how a variable handling subgraph is constructed in \cite{chung2009}.

From the sink node of the variable handling subgraph, we create three adapters to each of the interfaces corresponding to the clauses.  Each adapter corresponds to a literal in the clause, and the sole method in the interface is available only if the method corresponding to the literal is available.  And from each interface corresponding to a clause, there is a single adapter to the target interface for the entire graph which makes the method corresponding to the clause available only if the sole method in the clause interface is available.

For the graph constructed this way, the entire graph is obviously maximally covering with $2v+4c$ adapters and all methods available at the target interface.  If the original boolean expression is an instance of one-in-three 3SAT, then a satisfying assignment can specify a singly-linked path through the variable interfaces, followed by each true literal specifying the adapters to pass through to each clause interface, followed by the adapters to the target interface, and the resulting directed acyclic graph is a maximally covering web of adapters with $v+2c$ adapters, since all methods will be available at the target interface.

Conversely, if there is a maximally covering web of adapters with $v+2c$ adapters, then $2c$ adapters connect to the clause interfaces since all clause interfaces must be included.  The remaining $v$ adapters must be a singly-linked path through the variable interfaces, and the selection of adapters for each variable interface specifies a variable assignment which satisfies the original boolean expression with only one true literal in each clause.  Therefore MINWEB is NP-complete, and we can also conclude that minimizing the number of required adapters to adapt a single method is also NP-complete by removing the other methods in the target interface.

\section{Related work}
\label{sec:related-work}

\cite{gschwind:sem2002}  implements a network repository of interface adapters for adapting Java interfaces using single chains of adapters.  \cite{ponnekanti:percom2003} implements a similar adaptation framework for network services.  Although it allows an interface adapter to adapt a target interface from multiple source interfaces, only a single interface adapter is used for each target interface, so it has the same limitations as single adapter chains in that not all methods that could be adapted may actually be adapted.  Both mention the possibility of lossy interface adaptation, but neither considers how to minimize such loss.

\cite{kim:percom2008} proposes an interface adaptation framework which attempts to minimize the loss incurred by an interface adapter chain, and \cite{chung2009} rigorously defines the mathematical background required to implement such a framework.  These only consider the use of single chains of interface adapters.

\section{Conclusions}
\label{sec:conclusions}

We described a polynomial-time algorithm which can construct a maximally covering web of interface adapters, which may include a much larger number of interface adapters than necessary.  However, we also showed that minimizing the number of interface adapters included in a maximally covering web of interface adapters is an NP-complete problem.

Further work can be done to extend these results by relaxing the assumptions made for this paper.  We can consider the case when a method in a target interface can only be \emph{partially} implemented from methods in a source interface.  We can also consider how the quality of adapters should be dealt with in algorithms, or how to deal with adapters that maintain state.  Heuristic algorithms which attempt to minimize the number of adapters included in a nearly maximally covering web of interface adapters is another area for future work.

\bibliographystyle{abbrv}
\bibliography{strings,articles,hearsay,local,proceedings}

\end{document}